\documentclass[epj,nopacs]{svjour}

\usepackage{amssymb}
\usepackage{epsfig}
\usepackage{eucal}
\usepackage{amsmath}

\newcommand{\ta}[1]{#1\hspace{-.42em}/\hspace{-.07em}} 
\newcommand{\beq}{\begin{equation}}
\newcommand{\eeq}{\end{equation}}
\newcommand{\bea}{\begin{eqnarray}}
\newcommand{\eea}{\end{eqnarray}}
\newcommand{\non}{\nonumber}
\newcommand{\eq}[1]{eq.(\ref{#1})}         

\newcommand{\e}{\varepsilon} 
\newcommand{\be}{\beta} 

\title{NLO QED corrections to ISR in $e^+ e^-$ annihilation 
and the measurement of $\sigma(e^+ e^- \rightarrow hadrons)$ 
using tagged photons}

\author{Germ\'an Rodrigo 
\thanks{\email{rodrigo@particle.uni-karlsruhe.de}} \and
Aude Gehrmann-De Ridder 
\thanks{\email{gehra@particle.uni-karlsruhe.de}} \and 
Marc Guilleaume \thanks{\email{marc.giom@gmx.de}} \and 
Johann H. K\"uhn \thanks{\email{jk@particle.uni-karlsruhe.de}} }
\institute{Institut f\"ur Theoretische Teilchenphysik,
Universit\"at Karlsruhe, D-76128 Karlsruhe, Germany.}

\date{Received: June 13, 2001}

\abstract{The leptonic tensor for the process 
$e^+ e^- \rightarrow \gamma+\gamma^*$, which describes the
next-to-leading order virtual and soft QED corrections to
initial state radiation in $e^+ e^-$ annihilation with emission 
of an extra virtual photon decaying into hadrons, is calculated. 
A Monte Carlo generator for the reaction 
$e^+ e^- \rightarrow \gamma + \pi^+ \pi^-$ has
been set up which includes these corrections.
It thus describes configurations where the invariant mass of the
hadrons plus photon is very close to $\sqrt{s}$.
Predictions for cms energies of $1$ to $10$~GeV,
corresponding to the energies of DAPHNE and $B$-meson factories, 
are presented. The possibility for an accurate measurement 
using tagged photons of $\sigma(e^+ e^- \rightarrow hadrons)$, 
which plays an important role in the theoretical description of the 
muon anomalous magnetic moment and the running of the electromagnetic 
coupling, is discussed. }

\begin{document}

\authorrunning{G.~Rodrigo, A.~Gehrmann, M.~Guilleaume and J.H.~K\"uhn}
\titlerunning{NLO QED corrections to ISR in $e^+ e^-$ annihilation}
\maketitle


\section{Introduction}

Electroweak precision measurements have become one of the central
issues in present particle physics. The recent measurement
of the muon anomalous magnetic moment $(g-2)_{\mu}$ reported
in~\cite{Brown:2001mg} shows a discrepancy at the $2.6$ 
standard deviation level with respect to the theoretical evaluation 
of the same quantity~\cite{Hughes:1999fp}.
This disagreement, which has been taken as an indication of new physics,
has triggered a raving and somehow controversial deluge of 
non-Standard Model explanations.

One of the main ingredients of the theoretical prediction 
for the muon anomalous magnetic moment
is the hadronic vacuum polarization contribution~\cite{hadronicmuon}
which moreover is responsible for the bulk of the theoretical error.
It is in turn related via dispersion relations to the 
cross section for electron-positron annihilation into hadrons,
$\sigma_{had}$. This quantity plays also an important role in the 
evolution of the electromagnetic coupling $\alpha_{QED}$ from the 
low energy Thompson limit to high energies~\cite{hadronicmuon,runningQED}.
This indeed means that the interpretation of improved measurements 
at high energy colliders like LEP, LHC, or Tevatron depends 
significantly on the precise knowledge of $\sigma_{had}$.

The evaluation of the hadronic vacuum polarization
contribution to the muon anomalous magnetic moment, and even more so
to the running of $\alpha_{QED}$, requires the measurement of
$\sigma_{had}$ over a wide range of energies. Of particular 
importance is the low energy region around $2$~GeV, 
where $\sigma_{had}$ is at present experimentally poorly 
determined and only marginally consistent with the 
predictions based on pQCD. New efforts are therefore 
mandatory in this direction, which could help to either 
remove or sharpen the discrepancy between theoretical 
prediction and experimental results for $(g-2)_{\mu}$.

The feasibility of using tagged photons at high luminosity
electron-positron storage rings, like the $\phi$-factory DAPHNE
or at $B$-factories, to measure $\sigma_{had}$
has been proposed and studied in detail 
in~\cite{Binner:1999bt,Melnikov:2000gs} 
(See also~\cite{Spagnolo:1999mt,Khoze:2001fs}).
In this case, the machine is operating at a fixed cms energy and
initial state radiation is used to reduce the effective energy 
and thus the invariant mass of the hadronic system. The measurement 
of the tagged photon energy helps to constrain the kinematics. 
Preliminary experimental results using this method have been 
presented recently by the KLOE collaboration~\cite{Adinolfi:2000fv}.
Large event rates were also observed by the BaBar 
collaboration~\cite{babar}.

In this paper we consider the next-to-leading order (NLO) QED
corrections to initial state radiation (ISR) in the annihilation 
process $e^+ e^- \rightarrow \gamma + hadrons$.
After factorizing the hadronic contribution, the leptonic tensor, 
which contains virtual and soft photon corrections, is calculated. 
An improved Monte Carlo generator including these terms has been set-up. 
Predictions for the exclusive channel 
$e^+ e^- \rightarrow \pi^+ \pi^- \gamma$ at cms energies of 
$1$ to $10$~GeV, corresponding to the energies of DAPHNE and 
$B$-meson factories, are presented. 
The comparison with the EVA~\cite{Binner:1999bt} Monte Carlo,
simulating the same process at leading-order (LO), is performed.

\section{The leptonic tensor at leading order}

\begin{figure}
\begin{center}
\epsfig{file=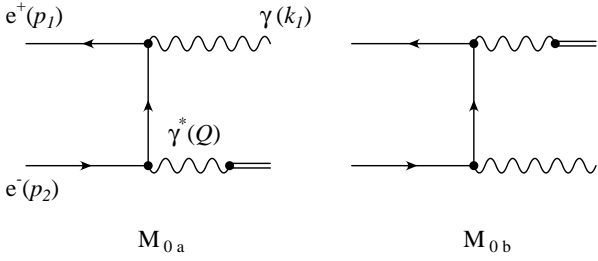,width=8cm}
\end{center}
\caption{Initial state radiation in the annihilation 
process $e^+ e^- \rightarrow \gamma + hadrons$ at the Born level.}
\label{fig:born}
\end{figure}

Consider the $e^+ e^-$ annihilation process
\begin{align}
e^+(p_1) + e^-(p_2) \rightarrow & \gamma^*(Q) + \gamma(k_1)~,
\non \\ & \; \; \; \downarrow \non \\ & \! \! hadrons
\end{align}
where the virtual photon decays into a hadronic final state and the 
real one is emitted from the initial state (figure~\ref{fig:born}).
The differential rate can be cast into the product of a leptonic 
and a hadronic tensor and the corresponding factorized phase space
\begin{equation}
d\sigma = \frac{1}{2s} L_{\mu \nu} H^{\mu \nu}
d \Phi_2(p_1,p_2;Q,k_1) d \Phi_n(Q;q_1,\cdot,q_n) 
\frac{dQ^2}{2\pi}~,
\end{equation}
where $d \Phi_n(Q;q_1,\cdot,q_n)$ denotes the hadronic 
$n$-body phase-space with all the statistical factors 
coming from the hadro\-nic final state included. 

For an arbitrary hadronic final state, the matrix element for 
the diagrams in figure~\ref{fig:born} is given by  
\begin{align}
M_0 &= M_0^{\mu} J_{\mu} = (M_{0a}^{\mu} + M_{0b}^{\mu}) J_{\mu} = \\
&= - \frac{e^2}{Q^2}  \bar{v}(p_1) \bigg(   
\frac{\ta{\varepsilon}^*(k_1)[\ta{k}_1-\ta{p}_1+m_e]
\gamma^{\mu}}{2 k_1 \cdot p_1} \non \\ 
& \qquad + \frac{\gamma^{\mu}[\ta{p}_2-\ta{k}_1+m_e]\ta{\varepsilon}^*(k_1)}
{2 k_1 \cdot p_2} 
\bigg) u(p_2) \; J_{\mu}~, \non 
\end{align}
where $J_{\mu}$ is the hadronic current.
Summing over the polarizations of the final real photon,
averaging over the polarizations of the initial $e^+ e^-$ state,
and using current conservation, $Q_{\mu} J^{\mu} = 0$, 
the leptonic tensor
\begin{equation*}
L_0^{\mu \nu} = \overline{M_0^{\mu} M_0^{\nu +}}~,
\end{equation*}
can be written in the following form: 
\begin{align}
L_0^{\mu \nu} &= 
\frac{(4 \pi \alpha/s)^2}{q^4} \; \bigg[ \left( 
\frac{2 m^2 q^2(1-q^2)^2}{y_1^2 y_2^2}
- \frac{2 q^2+y_1^2+y_2^2}{y_1 y_2} \right) g^{\mu \nu} \non \\ & 
+ \left(\frac{8 m^2}{y_2^2} - \frac{4q^2}{y_1 y_2} \right) 
\frac{p_1^{\mu} p_1^{\nu}}{s} 
+ \left(\frac{8 m^2}{y_1^2} - \frac{4q^2}{y_1 y_2} \right) 
\frac{p_2^{\mu} p_2^{\nu}}{s} \non \\
& - \left( \frac{8 m^2}{y_1 y_2} \right) 
\frac{p_1^{\mu} p_2^{\nu} + p_1^{\nu} p_2^{\mu}}{s} \bigg]~, 
\label{Lmunu0}
\end{align}
with 
\begin{equation}
y_i = \frac{2 k_1 \cdot p_i}{s}~, 
\qquad m^2=\frac{m_e^2}{s}~, \qquad q^2=\frac{Q^2}{s}~.
\label{dimensionless}
\end{equation}
It is symmetric under the exchange of the electron and 
the positron momenta. Expressing the bilinear products $y_i$ 
by the photon emission angle in the center of mass frame
\begin{equation*}
y_{1,2} = \frac{1-q^2}{2}(1 \mp \be \cos \theta)~, 
\qquad \be = \sqrt{1-4m^2}~,
\end{equation*}
and rewriting the two-body phase space  
\begin{equation}
d \Phi_2(p_1,p_2;Q,k_1) = \frac{1-q^2}{32 \pi^2} d \Omega~,
\end{equation}
it is evident that expression (\ref{Lmunu0}) contains several 
singularities: soft singularities for $q^2\rightarrow 1$ and 
collinear singularities for $\cos \theta \rightarrow \pm 1$.
The former are avoided by requiring a minimal photon energy.
The later are regulated by the electron mass.
For $s \gg m_e^2$, the expression (\ref{Lmunu0}) can be 
nevertheless safely taken in the limit $m_e\rightarrow 0$ if the 
emitted real photon lies far from the collinear region.
In general, however, one encounters spurious singularities in the 
phase space integrations if powers of $m^2=m_e^2/s$ are prematurely 
neglected.

The physics of the hadronic system, whose description is model dependent, 
enters only through the hadronic tensor 
\begin{equation}
H^{\mu \nu} = J^{\mu} J^{\nu +}~,
\end{equation}
where the hadronic current has to be parametrized through form factors.
For two charged pions in the final state, the current 
\begin{equation}
J^{\mu}_{2\pi} = i e F_{2\pi}(Q^2) \; (q_{\pi^+}-q_{\pi^-})^{\mu}~,
\end{equation}
where $q_{\pi^+}$ and $q_{\pi^-}$ are the momenta of the $\pi^+$ and
$\pi^-$ respectively, is determined by only one function,
the pion form factor $F_{2\pi}$ \cite{Kuhn:1990ad}.
The hadronic current for four pions exhibits a more complicated structure
and has been discussed in~\cite{Czyz:2000wh}.

After integrating the hadronic tensor over the hadronic 
phase space, one gets
\begin{equation}
\int H^{\mu \nu} \Phi_n(Q;q_1, \cdot, q_n) 
= \frac{e^2}{6\pi} (Q^{\mu}Q^{\nu} - g^{\mu \nu} Q^2) R(Q^2)~,
\end{equation}
where $R(Q^2) = \sigma(e^+ e^- \rightarrow hadrons)/
\sigma(e^+ e^- \rightarrow \mu^+ \mu^-)$.
After the additional integration over the photon angles,  
the differential distribution 
\begin{align}
Q^2 \frac{d\sigma}{dQ^2} = \frac{4\alpha^3}{3 s} R(Q^2)
\bigg\{ \frac{s^2+Q^4}{s(s-Q^2)} 
\left( \log \frac{s}{m_e^2}-1 \right)
\bigg\}~,
\label{diff1}
\end{align}
is obtained. 
If instead the photon polar angle is restricted to be 
in the range $\theta_{min}< \theta < \pi-\theta_{min}$,
this differential distribution is given by 
\begin{align}
Q^2 \frac{d\sigma}{dQ^2} &= \frac{4\alpha^3}{3 s} R(Q^2)
\bigg\{ \frac{s^2+Q^4}{s(s-Q^2)} 
\log \frac{1+\cos \theta_{min}}{1-\cos \theta_{min}} \non \\
& - \frac{s-Q^2}{s} \cos \theta_{min} \bigg\}~.
\label{diff2}
\end{align}
In the later case, the electron mass can be taken equal to zero
before integration, since the collinear region is excluded
by the angular cut. The contribution of the two pion exclusive
channel can be calculated from \eq{diff1} and \eq{diff2} by the substitution
$R(Q^2) \rightarrow (1-4m_{\pi}^2/Q^2)^{3/2}$ $\mid~F_{2\pi}(Q^2)\mid^2/4$.

\section{Virtual and soft photon corrections to the leptonic tensor}

\begin{figure}
\begin{center}
\epsfig{file=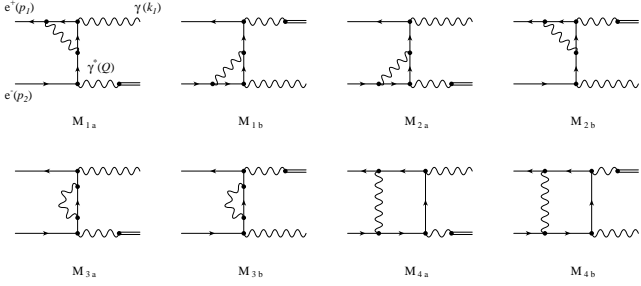,width=8.5cm}
\end{center}
\caption{One-loop corrections to initial state radiation in 
$e^+ e^- \rightarrow \gamma + hadrons$.}
\label{fig:nlo}
\end{figure}

At NLO, the leptonic tensor receives contributions 
both from one-loop corrections (figure~\ref{fig:nlo})
arising from the insertion of virtual photon lines in the tree 
diagrams of figure~\ref{fig:born} and from
the emission of an extra real photon from the initial state. 
In this paper, we consider only the emission of soft photons. 
The contribution of a second hard photon will be considered 
in a separate work~\cite{inpreparation}.

The one-loop matrix elements contribute to the leptonic tensor
through their interference with the lowest order diagrams
(figure~\ref{fig:born}).
They contain ultraviolet (UV) and infrared (IR) divergences
which are regularized using Dimensional Regularization
in $D=4-2\e$ dimensions.
The UV divergences are renormalized in the on-shell scheme.
The IR divergences are canceled by adding the contribution
of an extra soft photon emitted from the initial state
and integrated in the phase space up to an energy cutoff
$E_{\gamma}<w\sqrt{s}$ far below $\sqrt{s}$. The result, 
which is finite, depends, however, on this soft photon cutoff.
Only the contribution from hard photons with energy 
$E_{\gamma}>w\sqrt{s}$ would cancel this dependence.

The algebraic manipulations have been carried out with the 
help of the {\it FeynCalc} Mathematica package~\cite{Mertig:1991an}. 
Using standard techniques~\cite{Passarino:1979jh}, it
automatically reduces the evaluation of the one-loop contribution to 
the calculation of a few scalar one-loop integrals, listed in 
appendix~\ref{loopintegrals}, and performs the Dirac algebra. 

At NLO, the leptonic tensor has the following general form: 
\begin{align}
L^{\mu \nu} &=  \frac{(4 \pi \alpha/s)^2}{q^4 \; y_1 \; y_2} \;
\bigg[ a_{00} \; g^{\mu \nu} + a_{11} \; \frac{p_1^{\mu} p_1^{\nu}}{s}
 + a_{22} \; \frac{p_2^{\mu} p_2^{\nu}}{s} \non \\
&+ a_{12} \; \frac{p_1^{\mu} p_2^{\nu} + p_2^{\mu} p_1^{\nu}}{s}
+ i \pi \; a_{-1} \; 
\frac{p_1^{\mu} p_2^{\nu} - p_2^{\mu} p_1^{\nu}}{s} \bigg]~. 
\label{generaltensor}
\end{align}
The scalar coefficients $a_{ij}$ and $a_{-1}$ allow the following expansion 
\begin{equation}
a_{ij} = a_{ij}^{(0)} + \frac{\alpha}{\pi} \; a_{ij}^{(1)}~, \qquad
a_{-1} = \frac{\alpha}{\pi} \; a_{-1}^{(1)}~.
\end{equation}
The LO coefficients $a_{ij}^{(0)}$ can be directly read from~\eq{Lmunu0}.
For vanishing electron mass 
\begin{align}
a_{00}^{(0)} &= -(2q^2+y_1^2+y_2^2)~,  \qquad 
a_{12}^{(0)} = 0~, \non \\  
a_{11}^{(0)} &= a_{22}^{(0)} = -4q^2~.
\end{align}
Below, we use $m_e^2 \ll s, Q^2$ and neglect terms proportional
to $m_e^2$, which is a valid approximation if the observed
photon is far from the collinear region. The imaginary antisymmetric
piece proportional to $a_{-1}$ appears for the first time
at the NLO. The leptonic tensor remains therefore fully symmetric 
only at the LO. 

After adding the one-loop and the soft contribution, we 
end up with expressions for the NLO coefficients $a_{ij}^{(1)}$
and $a_{-1}^{(1)}$ which can be found in appendix~\ref{tensorcoefficients}.
As a test of our calculation, the leptonic tensor has been contracted 
with $(q^{\mu}q^{\nu}-q^2 g_{\mu\nu})$. The result:
\begin{align}
L^{\mu\nu} & (q_{\mu}q_{\nu}-q^2 g_{\mu\nu}) = 
L_0^{\mu\nu} (q_{\mu}q_{\nu}-q^2 g_{\mu\nu}) \\ \times &\bigg\{1 +
\frac{\alpha}{\pi} \bigg[ -\log(4w^2) [1+\log(m^2)] \non \\ 
& - \frac{3}{2} \log(\frac{m^2}{q^2}) - 2 + \frac{\pi^2}{3} 
+ \frac{2y_1 y_2}{2q^2+y_1^2+y_2^2} \non \\ & \times \bigg( 
\frac{1+(1-y_2)^2}{2y_1 y_2} L(y_1) 
+ \frac{1-2q^2}{2(1-q^2)^2} \log(q^2) \non \\
&+ \bigg[ 1-\frac{y_1-2y_2}{2(1-y_2)} - \frac{y_1 y_2}{4(1-y_2)^2}
\bigg] \log(\frac{y_1}{q^2}) \non \\
& - \frac{1}{2(1-q^2)} + \frac{1}{4(1-y_1)} + 
\frac{1-y_2}{4y_1} + (y_1 \leftrightarrow y_2) 
\bigg) \bigg] \bigg\}~, \non 
\end{align}
where
\begin{align}
L_0^{\mu\nu} (q_{\mu}q_{\nu}-q^2 g_{\mu\nu}) 
= (4 \pi \alpha/s)^2 \; \frac{2(2q^2+y_1^2+y_2^2)}{q^2 y_1 y_2}~,
\end{align}
reproduces, up to coupling constants and global factors, the one 
given in~\cite{Berends:1986yy} for the QED radiative corrections 
to the reaction $e^+ e^- \rightarrow Z \gamma$ .

\begin{figure}
\begin{center}
\epsfig{file=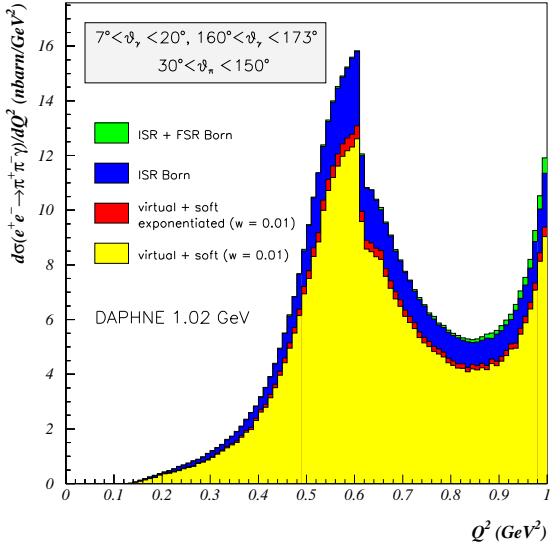,width=8cm}
\end{center}
\caption{Initial state radiation in the process
$e^+ e^- \rightarrow \pi^+ \pi^- \gamma$ at the Born (blue) and NLO level 
for $\sqrt{s}=1.02$~GeV. Virtual plus soft contributions are considered 
with a soft cutoff equal to $w=0.01$ at fixed order (yellow) and 
exponentiated (red). The final state radiation contribution 
(including interference with ISR) is also shown (green). 
The cuts are: $7^o<\theta_{\gamma}<20^o$ or $160^o<\theta_{\gamma}<173^o$,
$30^o<\theta_{\pi}<150^o$ and the energy of the observed photon
$E_{\gamma}>0.02$~GeV.}
\label{fig:nlosoft}
\end{figure}

\begin{figure}
\begin{center}
\epsfig{file=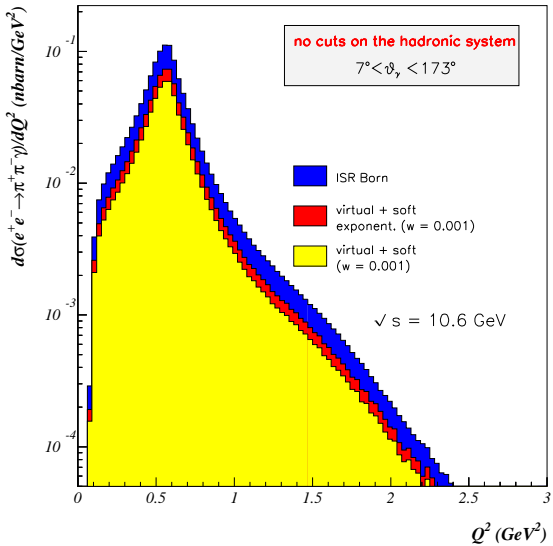,width=8cm}
\end{center}
\caption{Initial state radiation in the process
$e^+ e^- \rightarrow \pi^+ \pi^- \gamma$ at the Born (blue) and NLO level 
for $\sqrt{s}=10.6$~GeV. Virtual plus soft contributions are considered 
with a soft cutoff equal to $w=0.001$ at fixed order (yellow) and 
exponentiated (red). The cuts are: $7^o<\theta_{\gamma}<173^o$ 
and the energy of the observed photon $E_{\gamma}>0.02$~GeV.}
\label{fig:nlosoftb}
\end{figure}

The leptonic tensor can be cast, from~\eq{generaltensor}
and appendix~\ref{tensorcoefficients}, into the form:
\begin{align}
L^{\mu\nu} &= L_0^{\mu\nu} \bigg\{1 +
\frac{\alpha}{\pi}  \bigg[ -\log(4w^2) [1+\log(m^2)]  \non \\ 
&- \frac{3}{2} \log(m^2) - 2 + \frac{\pi^2}{3} \bigg] \bigg\}
+ C^{\mu\nu}~, 
\label{nloleptonic}
\end{align}
where the expected soft and collinear behaviour is manifest.
The first term, $\log(4w^2) [1+\log(m^2)]$, coming from the emission of
soft photons, gives a large contribution which eventually could spoil the
improvement expected from a NLO analysis.
The soft cutoff $w$ should be small enough to justify the soft
photon approximation. On the other hand, a very small value of $w$
would result in a meaningless negative cross-section.
To overcome this difficulty, the contribution of a second hard photon,
with energy $E_{\gamma}>w \sqrt{s}$, should be added~\cite{inpreparation},
thus canceling the $w$-dependence.
However, one may also limit the analysis to configurations where
the additional radiated photon is required to be soft,
by constraining the invariant mass of the hadron + photon system.
For small $w$ the following exponentiation can be used: 
\begin{align}
1 - \frac{\alpha}{\pi} \log(4w^2) [1+\log(m^2)] 
\rightarrow (2w)^{-\frac{2\alpha}{\pi} [1+\log(m^2)]}~.
\label{expo}
\end{align}
This accounts for the leading soft logarithms to all orders in 
perturbation theory~\cite{exponentiation} and vanishes 
in the limit $w\rightarrow 0$, as expected.

Another contribution which can be rather large and was
not considered up to now, is the insertion of vacuum polarization
corrections to the virtual photon line in figure~\ref{fig:born}.
Its inclusion does not affect the other features of our
calculation. It introduces a correction which is proportional
to the Born leptonic tensor and can be reabsorbed in the
choice of a running coupling constant.
In order to take into account also the higher orders, a factor
$1/(1-\delta_{VP})$ where $\delta_{VP}$ is the vacuum polarization 
contribution can be included. In the present version of this 
program, this factor has been dropped for simplicity. 

Motivated by these considerations, the following improved leptonic 
tensor can be defined:
\begin{align}
\label{improved}
L^{\mu\nu} &=  
\frac{(2w)^{-\frac{2\alpha}{\pi} [1+\log(m^2)]}}{1-\delta_{VP}} \\ \times &
\bigg[ L_0^{\mu\nu} \bigg\{1 + \frac{\alpha}{\pi}  \bigg[ 
- \frac{3}{2} \log(m^2) - 2 + \frac{\pi^2}{3} \bigg] \bigg\}
+ C^{\mu\nu} \bigg]~.  \non 
\end{align}

\section{Monte Carlo simulation}

A Monte Carlo generator has been built. It simulates the
process $e^+e^- \rightarrow \pi^+ \pi^- \gamma$ where the
photon is emitted from the initial state, with a large 
angle with respect to the beam direction. It is based
on EVA~\cite{Binner:1999bt} and includes the NLO corrections
described in the previous sections~\footnote{The default version 
of the program is based on eqs(\ref{gmunu}-\ref{asym}) and is 
thus valid in the limit $m_e^2/s \ll 1$. As an alternative, it is 
also possible to run the program with formulae which include 
the complete $m_e$ dependence.}. The program is built in 
a modular way such that the simulation of other exclusive 
hadronic channels can be easily included while keeping 
the factorization of the initial state QED corrections.
Configurations with an additional hard photon are not yet included.
Hence the generator can only be used to describe configurations
where the invariant mass of the hadronic system plus the tagged
photon is close to the total energy of the collision,
$M(\gamma+hadrons) > (1-w)\sqrt{s}$. A generator which 
includes also emission of two hard photons will be presented in 
a subsequent publication. 

In figure~\ref{fig:nlosoft}, the differential distribution
in the invariant mass of the hadronic system is shown for 
the process $e^+e^- \rightarrow \pi^+ \pi^- \gamma$
at DAPHNE energies, $\sqrt{s}=1.02$~GeV.
In principle, initial and final state radiation (FSR) would be 
required for the complete simulation of the process.
However, the cuts: pions in the central region, $30^o<\theta_{\pi}<150^o$,
and photons close to the beam and well separated from the pions,
$7^o<\theta_{\gamma}<20^o$ or $160^o<\theta_{\gamma}<173^o$,
select configurations dominated by
initial state radiation~\cite{Binner:1999bt,Melnikov:2000gs}.
For comparison, the contribution of the final state radiation
at Born level is also shown. 
The corrections to the Born approximation reach up to $20\%$
for a soft cutoff of $w=0.01$ if the fixed order expression
(\ref{nloleptonic}) is used, or $17\%$ if the 
exponentiation of the soft cutoff, \eq{expo}, is applied.

At B-factories, $\sqrt{s}=10.6$~GeV, the situation is slightly different. 
Because of the higher center of mass energy, the hadronic system and the
real photon are mainly produced back to back. 
The FSR is therefore already suppressed without imposing additional 
kinematical cuts on the hadronic system. 
For these high energies one might in addition include 
photons emitted also at larger angles. From the analysis of 
EVA~\cite{Binner:1999bt} one finds that the events are always 
dominated by ISR -- a consequence of the strong form factor
suppression of FSR at high energies and large angles between 
pions and photons. Figure~\ref{fig:nlosoftb} shows the differential 
distribution in the invariant mass of the hadronic system 
at $\sqrt{s}=10.6$~GeV for photons in the region 
$7^o<\theta_{\gamma}<173^o$ and no constraints to the hadronic 
system. The FSR amounts to less than $10^{-3}$ and is not shown.
The NLO corrections are again dominated by the value of the 
soft photon cutoff.

\section{Leading log resummation}

From the results of the previous section, it is clear that the
fixed order correction is dominated by the value of the soft photon 
cutoff $w$.
Even if the exponentiation is applied, which accounts for the leading 
soft logarithms to all orders in perturbation theory, the situation 
does not improve much.
The addition of the contribution of a second hard photon would
cancel the strong $w$-dependence, however, large logarithms
of collinear origin, $L=\log(s/m_e^2)$, would remain.
Because these large logarithms will show up in all orders of
perturbation theory, resummation techniques have been developed
which are constructed to include the dominant higher order terms.
The method of structure
functions~\cite{structure,Berends:1988ab} accounts well for these
corrections. It takes into account all the leading logarithmic
corrections $\cal{O}$$(\alpha L)$, coming from virtual, soft, and 
hard photon contributions to all orders in perturbation theory.

\begin{figure}
\begin{center}
\epsfig{file=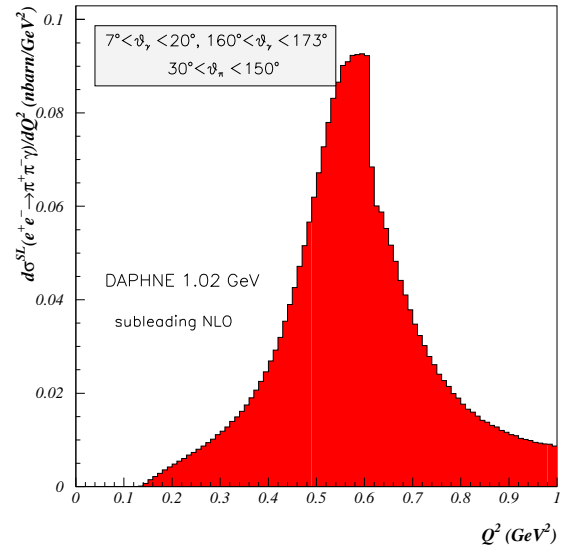,width=8cm}
\end{center}
\caption{Subleading NLO contribution to 
initial state radiation in the process
$e^+ e^- \rightarrow \pi^+ \pi^- \gamma$.
Same cuts as in figure~\ref{fig:nlosoft}.}
\label{fig:subleading}
\end{figure}

In~\cite{Binner:1999bt} these effects were considered in the simulation
of the process $e^+ e^- \rightarrow  \pi^+ \pi^- \gamma$.
The formulas of~\cite{Caffo:1997yy} were used which include terms
up to order $\alpha^2 L^2$. A minimal invariant mass of the
$\pi^+ \pi^- \gamma$ system was required in order to reduce the
kinematic distorsion of the events due to the collinear initial
state radiation. For a sufficiently tight cut this corresponds to
the situation discussed in the present paper. For a minimal invariant
mass of the $\pi^+ \pi^- \gamma$ system of $0.9$~GeV$^2$, a negative 
$6\%$ correction to the Born approximation was found. 

Subleading effects, which correspond to the $C^{\mu \nu}$
terms of our leptonic tensor in~\eq{nloleptonic},
are not taken into account within this approximation.
In figure~\ref{fig:subleading}, the contribution of these subleading
terms to the differential distribution of figure~\ref{fig:nlosoft}
is shown. They amount to $0.6\%$ of the Born approximation close to 
the $\rho$-resonance, where the rate is maximal, and to $1\%$ for small 
values of $Q^2$, where, however, the cross section rate is small. 
Similar values are found for higher center of mass energies. 
The matching of the resummed result, through the structure function 
technique, with the fixed order result 
is under study~\cite{inpreparation}.

\section{Conclusions}

ISR  at high luminosity $e^+ e^-$ colliders ($\phi$ and $B$-factories) 
is an alternative to the direct measurement of 
$\sigma(e^+e^-\rightarrow hadrons)$ giving access to a wide range of 
energies, from threshold to the center of mass energy of the collider.
The NLO QED correction to ISR, in the form of a leptonic tensor, 
has been calculated and included in a Monte Carlo generator, 
which was compared with EVA~\cite{Binner:1999bt} for the 
$\pi^+\pi^-$ exclusive channel.
The modular structure  of the calculation, independent of the final 
state hadronic system, is such that other hadronic channels or improved 
parametrizations of the hadronic current can be easily included.

From our results it is clear that the fixed order correction
is dominated by the value of the soft photon cutoff $w$.
Only the contribution of a second hard photon~\cite{inpreparation}
would cancel this strong $w$-dependence.
But even if this contribution is added, large logarithms
of collinear origin, $L=\log(s/m_e^2)$, would still remain.
Furthermore, they  will show up in all orders of
perturbation theory. A consistent way to resum 
such leading log terms together with subleading effects
is also under consideration~\cite{inpreparation}. 

\section*{Acknowledgments}

We would like to thank G.~Cataldi, A.~Denig, S.~Di~Falco, W.~Kluge, 
S.~M\"uller, G.~Venanzoni and B.~Valeriani for reminding us constantly 
of the importance of this work for the experimental analysis.
Work supported in part by BMBF under grant number 05HT9VKB0 and 
E.U. EURODAPHNE network TMR project FMRX-CT98-0169.

\appendix

\section{Tensor coefficients}
\label{tensorcoefficients}

In this appendix, we collect our results for the scalar coefficients
of the leptonic tensor~(\ref{generaltensor}) at the NLO
arising from virtual and soft photon contributions.
Our expressions are valid in the limit of small electron mass. 

For the coefficient proportional to $g^{\mu \nu}$, we get
\begin{align}
a_{00}^{(1)} &= 
  a_{00}^{(0)} \bigg[ -\log(4w^2) [1+\log(m^2)] \non \\
&-\frac{3}{2} \log(\frac{m^2}{q^2}) - 2 + \frac{\pi^2}{3} \bigg]  \non \\ 
&-\frac{q^2(1-q^2)}{2}  - y_1 y_2
 - \bigg[ q^2 + \frac{2y_1 y_2}{1-q^2} \bigg] \log(q^2) \non \\
&+ \frac{y_1}{2} \bigg[ 4-y_1-\frac{3(1+q^2)}{1-y_2} \bigg] 
\log(\frac{y_1}{q^2}) \non \\
&+ \frac{y_2}{2} \bigg[ 4-y_2-\frac{3(1+q^2)}{1-y_1} \bigg] 
\log(\frac{y_2}{q^2}) \non \\
&- \bigg[ 1+(1-y_2)^2+\frac{y_1 q^2}{y_2} \bigg] L(y_1) \non \\
&- \bigg[ 1+(1-y_1)^2+\frac{y_2 q^2}{y_1} \bigg] L(y_2)~,
\label{gmunu}
\end{align}
with 
\begin{align*}
L(y_i) = Li_2(\frac{-y_i}{q^2})-Li_2(1-\frac{1}{q^2}) + 
\log(q^2+y_i) \log(\frac{y_i}{q^2})~, 
\end{align*}
where $Li_2$ is the Spence or dilogarithm function and 
$y_i$, $q^2$ and $m^2$ have been defined in~\eq{dimensionless}.
The quantity $w$ denotes the dimensionless 
value of the soft photon energy cutoff: $E_{\gamma}<w\sqrt{s}$.
The coefficient in front of the tensor structure 
$p_1^{\mu}p_1^{\nu}$, is given by 
\begin{align}
\label{p1p1}
a_{11}^{(1)} &= 
  a_{11}^{(0)} \bigg[ -\log(4w^2) [1+\log(m^2)]  \\ 
&-\frac{3}{2} \log(\frac{m^2}{q^2}) - 2 + \frac{\pi^2}{3} \bigg]  \non \\ 
&+ (1+q^2)^2 \left(\frac{1}{1-y_1}-\frac{1}{1-q^2} \right) -
\frac{4(1-y_2)y_1}{1-q^2} \non \\
&- \frac{2 q^2}{1-q^2} \bigg[(1-y_2)\left(\frac{1}{y_2}+\frac{q^2}{y_1}
+ \frac{2 y_1}{1-q^2} \right) \non \\ & +\frac{2q^2}{1-q^2}
\bigg] \log(q^2) 
- q^2 \bigg[ 1 + \frac{2}{y_2} \bigg] \log(\frac{y_1}{q^2}) \non \\ 
&- q^2 \bigg[ \frac{(2-3y_1)(1-y_2)^2}{y_1(1-y_1)^2} \bigg] 
 \log(\frac{y_2}{q^2}) \non \\
&- 2 q^2 \bigg[ 1+\frac{1}{y_2^2} \bigg] L(y_1)
 - 2 q^2 \bigg[ 3+\frac{2q^2}{y_1}+\frac{q^4}{y_1^2} \bigg] L(y_2)~, \non 
\end{align}
that in front of $p_2^{\mu}p_2^{\nu}$ can be obtained by symmetric 
considerations, exchanging the positron with the electron momenta
\begin{align}
a_{22}^{(1)} = a_{11}^{(1)} (y_1 \leftrightarrow y_2)~.
\label{p2p2}
\end{align}
For the symmetric tensor structure 
($p_1^{\mu}p_2^{\nu}+p_2^{\mu}p_1^{\nu}$), we get 
\begin{align}
\label{p1p2}
a_{12}^{(1)} &= \frac{q^2}{1-y_1}+\frac{q^2}{1-y_2}
- \frac{4q^2+(y_1-y_2)^2}{1-q^2} \\
&-2 q^2 \bigg[ \frac{q^2}{y_1 y_2}+\frac{1+q^2-2 y_1 y_2}{(1-q^2)^2} \bigg] 
\log(q^2) \non \\
& - \frac{2 q^2}{1-y_2} \bigg[1-y_1+\frac{q^2}{y_2}-\frac{q^2}{2(1-y_2)} 
\bigg] \log(\frac{y_1}{q^2}) \non \\
& - \frac{2 q^2}{1-y_1} \bigg[1-y_2+\frac{q^2}{y_1}-\frac{q^2}{2(1-y_1)} 
\bigg] \log(\frac{y_2}{q^2}) \non \\
&- 2 q^2 \bigg[ 1+\frac{q^2}{y_2}+\frac{q^2}{y_2^2} \bigg] L(y_1)
 - 2 q^2 \bigg[ 1+\frac{q^2}{y_1}+\frac{q^2}{y_1^2} \bigg] L(y_2)~. \non 
\end{align}
Finally, the antisymmetric coefficient $a_{-1}$,
accompanying ($p_1^{\mu}p_2^{\nu}-p_2^{\mu}p_1^{\nu}$), reads 
\begin{align}
a_{-1}^{(1)} &= q^2 \bigg[ \frac{\log(m^2 y_2)}{y_1}  
+ \frac{1-q^2}{1-y_1} + \frac{q^2}{(1-y_1)^2} \bigg] \non \\
&- (y_1 \leftrightarrow y_2)~.
\label{asym}
\end{align}

\section{Loop integrals and soft photon contribution}
\label{loopintegrals}

The Passarino-Veltman procedure~\cite{Passarino:1979jh} allows to reduce
the calculation of any one-loop amplitude to the evaluation of a few 
scalar one-loop integrals. In this appendix, we collect the scalar 
one-loop integrals needed for our calculation and give their 
expression in the limit of small electron mass.
Ultraviolet (UV) and infrared (IR) divergences appear in the 
one-loop calculation. Dimensional regularization in $D=4-2\e$ 
dimensions is used to regularize both kind of divergences. 
The UV divergences are renormalized using the on-shell 
renormalization scheme. The soft photon contribution to the 
leptonic tensor cancels the remaining IR divergences.

A few two-, three-, and four-point scalar one-loop integrals
enter our calculation. Expression for the two-point scalar 
integrals are simple and well known.
The general three-point scalar one-loop integral is defined by 
\begin{align}
& C_0(p_a^2,(p_a-p_b)^2,p_b^2,m_1^2,m_2^2,m_3^2) 
= -i 16\pi^2 \mu^{4-D} \non \\ 
& \times \int \frac{d^D k}{(2\pi)^D} 
\frac{1}{[k^2-m_1^2][(k-p_a)^2-m_2^2][(k-p_b)^2-m_3^2]}~.  
\end{align}
Four different three-point scalar one-loop integrals are needed 
\begin{align}
& C01 = C_0((p_i-k_1)^2,0,m_e^2,0,m_e^2,m_e^2)~, \non \\ 
& C02 = C_0(m_e^2,s,m_e^2,0,m_e^2,m_e^2)~, \non \\
& C03 = C_0((p_i-k_1)^2,Q^2,m_e^2,0,m_e^2,m_e^2)~, \non \\
& C04 = C_0(Q^2,s,0,m_e^2,m_e^2,m_e^2)~,
\end{align}
$i=1,2$, and one scalar box 
\begin{align}
& D0  = -i 16\pi^2 \mu^{4-D} \int \frac{d^D k}{(2\pi)^D}  \\
& \times
\frac{1}{k^2[(k+p_i)^2-m_e^2][(k+p_i-k_1)^2-m_e^2][(k-p_j)^2-m_e^2]}~, \non    
\end{align}
with $j \ne i$. 
 
In the limit of vanishing electron mass: $m^2 \ll s, Q^2, y_i$, the 
following results are obtained:
\begin{align}
& C01 = 
\frac{s^{-1}}{y_i} \bigg[-\frac{1}{2} \log^2 \frac{m^2}{y_i} - \frac{\pi^2}{3}
\bigg]~, \non \\
& C02 = s^{-1} \bigg[ 
\log(m^2) \Delta - \frac{\log^2(m^2)}{2} - \frac{2\pi^2}{3}
+ i \pi \Delta \bigg]~, \non \\
& C03 =  \frac{s^{-1}}{q^2+y_i} \bigg[ 
\log(m^2) \log(\frac{y_i}{q^2}) - 2 \log (q^2+y_i) \log(\frac{y_i}{q^2})
+ \non \\ & \quad \frac{\log^2(y_i)}{2} - \frac{\log^2(q^2)}{2} 
- 2 Li_2(\frac{-y_i}{q^2}) - \frac{7\pi^2}{6}
+ i \pi \log(\frac{y_i}{q^2}) \bigg]~, \non \\
& C04 = \frac{s^{-1}}{1-q^2} \bigg[\log(m^2) - \frac{\log(q^2)}{2}  
+ i \pi \bigg] \log(q^2)~, \non \\
& D0 =  \frac{s^{-2}}{y_i}
\bigg[ - \log(m^2) \Delta  
+ 2 \log(m^2) \log(\frac{y_i}{q^2}) - 2 Li_2(1-\frac{1}{q^2}) \non \\
& \quad -\frac{5\pi^2}{6} - i \pi 
\bigg( \Delta-\log(m^2) - 2 \log(\frac{y_i}{q^2}) \bigg)
\bigg]~,
\end{align}
where 
\begin{equation}
\Delta = \frac{(4\pi)^\e}{\e \; \Gamma(1-\e)} 
\left( \frac{\mu^2}{s} \right)^{\e}~.
\label{delta}
\end{equation}
They can be compared, for instance, with the one-loop integrals
quoted in~\cite{Berends:1986yy} where a fake photon mass $\lambda$
was used to regularize the IR divergences. The identification 
$\Delta \leftrightarrow \log(\lambda^2/s)$ allows to pass from one 
scheme to the other.

\subsection{Initial state soft photon emission}

The contribution to the leptonic tensor of a soft photon with
momentum $k$ and energy $E<w \sqrt{s}$, with $w \ll 1$, reads 
\begin{align}
L_{soft}^{\mu \nu} &= -4 e^2 \frac{\mu^{4-D}}{2(2\pi)^{D-1}} 
\int_0^{w \sqrt{s}} E^{D-3} dE d\Omega \non \\ & \times 
\left( \frac{p_1}{2p_1\cdot k} - \frac{p_2}{2p_2\cdot k}\right)^2
L_0^{\mu \nu}~.
\end{align}
Integrated in the soft photon phase space~\cite{Rodrigo:1999gv} 
and in the limit of small electron mass, we get
\begin{align}
L_{soft}^{\mu \nu} = \frac{\alpha}{\pi} \bigg\{ &
\left[ \Delta - \log(4w^2) \right] \left[ 1+\log(m^2)\right] \non \\
& - \frac{\log^2(m^2)}{2} - \log(m^2) - \frac{\pi^2}{3} \bigg\}
L_0^{\mu \nu}~.
\label{Lsoft}
\end{align}
After UV renormalization, the renormalized one-loop matrix elements 
$\tilde{M}_i$ of the diagrams in figure~\ref{fig:nlo}, which contribute at 
the NLO to the leptonic tensor through their interference with the 
Born diagrams, give the following IR contribution  
\begin{align}
L_{IR}^{\mu \nu} &= \sum_{i=1,4} 
(\overline{\tilde{M}_i^{\mu} M_0^{\nu +} + M_0^{\mu} \tilde{M}_i^{\nu +}}
)\big|_{\e} = \non \\ & =
- \frac{\alpha}{\pi} 
\Delta \left[\log(m^2) + 1 \right]  L_0^{\mu \nu}~.
\label{LIR}
\end{align}
Summing up~\eq{Lsoft} and~\eq{LIR}, a finite result is found.


\end{document}